# Gate-controlled proximity magnetoresistance in $In_{1-x}Ga_xAs/(Ga,Fe)Sb$ bilayer heterostructures

Kosuke Takiguchi[1], Kyosuke Okamura[1], Le Duc Anh[1,2,3], and Masaaki Tanaka[1,4,5]

[1] *Department of Electrical Engineering and Information Systems, The University of Tokyo, Bunkyo-ku, Tokyo 113-8656, Japan*

[2] *Institute of Engineering Innovation, The University of Tokyo, Bunkyo-ku, Tokyo 113-8656, Japan.*

[3] *PRESTO, Japan Science and Technology Agency, Kawaguchi, Saitama, 332-0012, Japan*

[4] *Centre for Spintronics Research Network (CSRN), The University of Tokyo, Bunkyo-ku, Tokyo 113-8656, Japan.*

[5] *Institute for Nano Quantum Information Electronics, The University of Tokyo, 4-6-1, Komaba, Meguro-ku, Tokyo 153-8505 Japan.*

Abstract

The magnetic proximity effect (MPE), ferromagnetic coupling at the interface of magnetically dissimilar layers, attracts much attention as a promising pathway for introducing ferromagnetism into a high-mobility non-magnetic conducting channel. Recently, our group found giant proximity magnetoresistance (PMR), which is caused by MPE at an interface between a non-magnetic semiconductor InAs quantum well (QW) layer and a ferromagnetic semiconductor (Ga,Fe)Sb layer. The MPE in the non-magnetic semiconductor can be modulated by applying a gate voltage and controlling the penetration of the electron wavefunction in the InAs QW into the neighboring insulating ferromagnetic (Ga,Fe)Sb layer. However, optimal conditions to obtain strong MPE at the InAs/(Ga,Fe)Sb interface have not been clarified. In this paper, we systematically investigate the PMR properties of $In_{1-x}Ga_xAs$ ($x$ = 0%, 5%, 7.5%, and 10%) / (Ga,Fe)Sb bilayer semiconductor heterostructures under a wide range of gate voltage. The inclusion of Ga alters the electronic structures of the InAs thin film, in particular changing the effective mass and the QW potential of electron carriers. Our experimental results and theoretical analysis of the PMR in these $In_{1-x}Ga_xAs$/(Ga,Fe)Sb heterostructures show that the MPE depends not only on the degree of penetration of the electron wavefunction into (Ga,Fe)Sb but also on the electron density. These findings help us to unveil the microscopic mechanism of MPE in semiconductor-based non-magnetic/ferromagnetic heterojunctions.

**I. Introduction**

Introducing a ferromagnetic coupling into a high-mobility semiconducting channel can lead to new types of spintronics devices with non-volatile and reconfigurable functions. One straightforward method towards this goal is doping magnetic impurities into a non-magnetic semiconductor, which led to the creation of ferromagnetic semiconductors (FMSs), such as (III,TM)V where TM is a transition metal element (Mn, Fe). However, carrier transport in FMSs is subject to frequent scatterings by the magnetic impurities, thus the carrier mobility is very low (~1 – 10 $cm^2$/Vs). For electronic device applications, it is important to obtain large magnetic responses in non-magnetic channels with high carrier mobility, minimizing scatterings by magnetic impurities. Utilizing a magnetic proximity effect (MPE), which is a magnetic coupling at the magnetically dissimilar layers, is one of the most promising pathways for this purpose [1–7]. In particular, bilayer systems consisting of a nonmagnetic (NM) conductive channel and a ferromagnetic (FM) insulator hold a magnetic coupling via MPE at the interface. In addition, since these bilayer systems have only one FM layer, the device fabrication is much simpler than that in the conventional spin-valve structures which contain FM/NM/FM tri-layer and more complicated multilayer structures [8–11]. Therefore, the magnetotransport phenomena induced by MPE in the FM/NM bilayers have been actively studied in metallic systems in recent years [12–18]. However, at these metallic NM/FM interfaces, the magnetoresistance (MR) magnitude is too small (< 0.1 – 1 %) for practical purposes, which is mainly due to the short-range nature of the magnetic coupling (< 1 nm).

In contrast, *semiconductor-based* NM/FM bilayer systems can overcome the problem of the metallic counterparts because the higher coherency and smaller concentration of carriers enhance the interfacial magnetic coupling range. In particular, when we prepare a NM semiconducting quantum well (QW) interfaced with an insulating FM layer, as shown in Fig. 1(a), even small penetration of two-dimensional carrier's wavefunction in the QW into the neighbouring FMS layer is enough to yield a strong MPE. This is because the two-dimensional carrier system feels the spin-carrier interactions occurring at the interface *as a whole*, due to their high coherency. Furthermore, the MPE can be modulated by an external gate voltage, which enhances the penetration by pushing the wavefunction towards the FM side. Therefore, the spin splitting energy induced by MPE can be controlled by external electrical means.

Recently, we demonstrated the abovementioned advantage of the semiconductor NM/FM bilayer systems using a NM InAs QW/ FM (Ga,Fe)Sb [19]. The InAs/(Ga,Fe)Sb bilayer heterostructures have distinct merits as follows: (i) (Ga,Fe)Sb is a *p*-type FMS with high Curie temperature over 300 K [20,21]. (ii) The lattice mismatch between InAs and (Ga,Fe)Sb is only of the order of 0.1%, which allows epitaxial growth of high-quality single-crystalline heterostructures. (iii) InAs/(Ga,Fe)Sb has a type-III band lineup, in which the bottom of the conduction band of InAs is lower than the top of the valence band of (Ga,Fe)Sb at the NM/FM interface. This leads to large penetration of the electron wavefunction of the InAs QW into the (Ga,Fe)Sb side. (iv) At low temperature (< 5 K), the resistivity of (Ga,Fe)Sb is two-orders of magnitude higher than that of the InAs QW, thus electron carriers mainly flow in the InAs QW. The strong MPE led to a discovery of a new proximity magnetoresistance (PMR), whose magnitude reaches 20% at 10 T, which is twenty-fold larger than that of metallic systems, and a large spontaneous spin-splitting energy (~3.8 meV) in the InAs QW. Moreover, we successfully enhanced the PMR by applying a gate voltage.

However, in order to fully control the properties of the InAs/(Ga,Fe)Sb heterostructures and to realize practical spintronic applications, it is essential to obtain deeper insights into the microscopic mechanism of the MPE. To quantitatively estimate the MPE and spontaneous spin splitting energy induced in the NM channel, which is practically important, there are four main parameters that can be controlled experimentally; the penetration $P$ of the electron wavefunction of the NM QW channel into the FM side, the electron concentration $n$, the barrier height $E_b$, and the effective mass $m^*$ of the electron carriers in the NM QW. These four parameters are closely correlated. As shown in Fig. 1(a), the wavefunction $\varphi(z)$ decays exponentially in the FM insulator; $\varphi(z) = |\varphi_0| \exp[-(z - z_0)/\lambda]$, where $|\varphi_0|$ is the amplitude of the wavefunction at the interface, $z$ is the axis in the growth direction, and $z_0$ is the position at the interface of NM/FM. Then, the penetration $P$ is given by $P = |\varphi_0|\lambda$. According to the WKB approximation for tunnelling phenomena, $\lambda \propto 1/\sqrt{E_b}$ and $|\varphi_0| \propto (m^*)^{-1/4}$. Thus, large $E_b$ and/or $m^*$ suppress $P$ and consequently the MPE. Larger $n$ also leads to heavier $m^*$ due to the non-parabolicity of the InAs conduction band and supresses the MPE [22,23]. On the other hand, an important question is whether larger $n$ can induce stronger MPE or not via enhancement of the interfacial $s$-$d$ exchange coupling between electron carriers and Fe spins in the FMS, a well-known effect in conventional carrier-induced FMSs [24]. Therefore, in order to obtain larger spin splitting energy $\Delta E$ via MPE, investigating how $\Delta E$ depends on these trade-off parameters is crucial for seeking the optimal conditions.

In this paper, we investigate the PMR phenomena in field-effect transistor (FET) structures of $In_{1-x}Ga_xAs$ ($x$ = 0%, 5%, 7.5%, and 10%)/(Ga,Fe)Sb bilayers while applying a top gate voltage $V_g$. Compared with the previous study of PMR in InAs/(Ga,Fe)Sb [19], Ga inclusion in the InAs channel (InGaAs channel) changes the $E_b$ and $m^*$, which affects the penetration $P$. By the analysis of the PMR observed in these four samples, we find that $\Delta E$ depends on $n$ and the momentum relaxation time $\tau$ of electron carriers. These results suggest that $\Delta E$ is almost proportional to $n$ in an accumulation state of the FET operation, which indicates that the spin splitting via MPE is induced not only by the penetration $P$ of the wavefunction but also by the large carrier density $n$.

**II. Experiment**

We grew heterostructures consisting of $In_{1-x}Ga_xAs$ (15 nm, $x$ = 0%, 5%, 7.5%, 10%)/(Ga,Fe)Sb (15 nm, Fe 20%)/AlSb (140 nm)/AlAs (10 nm)/GaAs (100 nm) on semi-insulating GaAs (001) substrates by molecular beam epitaxy (MBE) (Fig. 1(b)). The growth temperature was 550 °C for the GaAs and AlAs layers, 470 °C for the AlSb layer, 250 °C for the (Ga,Fe)Sb layer and the $In_{1-x}Ga_xAs$ layer. The Ga content (= $x$) was determined by the ratio of In and Ga fluxes, which were calibrated using reflection high-energy electron diffraction (RHEED) intensity oscillations. *In situ* RHEED patterns of the $In_{1-x}Ga_xAs$ (Fig.1(c)-(f)) and (Ga,Fe)Sb (Fig.1(g)-(j)) layers were bright and streaky, indicating good crystal quality and smooth surface during the growth.

We first patterned the samples into 100 μm (width) × 400 μm (length) Hall bars using standard photolithography and Ar ion milling, and then formed several electrodes (source S, drain D, and electrodes for transport measurements) using sputtering deposition and lift-off of an Au (50 nm)/Cr (5 nm) film. An insulating $Al_2O_3$ layer (~50 nm) was deposited as a gate insulator at 170 °C on the Hall bars using atomic layer deposition (ALD). It is known that depositing $Al_2O_3$ using ALD can decrease the interfacial state density at the $In_{1-x}Ga_xAs$/oxide interface [25,26]. Thus, it is expected that the modulation of the electrical properties by the gate voltage in our $Al_2O_3/In_{1-}$ $_x$Ga$_x$As/(Ga,Fe)Sb FETs is more effective than that of the HfO$_2$/InAs/(Ga,Fe)Sb FETs reported in our previous study of PMR [19]. Finally, we formed a top-gate electrode (G), again by sputtering deposition and lift-off processes of an Au (50 nm)/Cr (5 nm) film. Figure 1(k) shows an optical microscope image of the FET device examined in this study. We applied a gate voltage $V_g$ between the G electrode and the S electrode. We measured the longitudinal resistance $R_{xx}$ by a standard four-terminal method and the Hall resistance $R_{xy}$ simultaneously at 2 K. In this study, a magnetic field $B$ was always applied perpendicular to the film plane. In our recent study on the magnetotransport properties of InAs/(Ga,Fe)Sb bilayers, we have found a new type of odd-parity magnetoresistance (OMR), which is an odd function against $B$, and the resistance change reaches almost 13.5% of the total resistance at 10 T [27]. Although the OMR is also observed in this study, this phenomena is out of the scope of this paper and will be discussed elsewhere. Therefore, all the MR and the Hall resistance data in this work were obtained by extracting only the even- and odd-function components against $B$ from the raw data, respectively. To obtain the QW potential and electron carrier wavefunctions in the In$_{1-x}$Ga$_x$As channel, we performed a self-consistent calculation for the case of $x = 5\%$ using Nextnano3. We note that the Fermi level in this calculation was assumed to be pinned at 0.60 eV below the conduction band bottom of (Ga,Fe)Sb due to the Fe impurity band in the band gap of this FMS as estimated in our previous study [28].

**III. RESULTS & DISCUSSIONS**

Figure 2(a)-(d) show the MR data of In$_{1-x}$Ga$_x$As ($x$ = 0%, 5%, 7.5% and 10%)/(Ga,Fe)Sb FET devices, named as D$_0$, D$_5$, D$_{7.5}$, and D$_{10}$, respectively, with a wide range of $V_g$ (−10 V < $V_g$ < 10 V). As summarized in Fig. 3(a), the MRs at 1 T (= [$R_{xx}$($B$ =

1 T)−$R_{xx}$(0 T)]/$R_{xx}$(0 T))) in $D_0$, $D_5$, and $D_{10}$ show a systematic change from negative to positive when negative $V_g$ is applied from 0 to −10 V. In $D_{7.5}$, the negative MR is suppressed to nearly zero at $V_g$ = −8 V. This variation of PMR can be understood by contributions of both Kondo scattering and s-d exchange interaction at the NM/FM interface, as will be explained later. We fit a modified Khosla-Fischer model [19,29] using the following equations (eqs. (1) – (5)) to our experimental results shown in Fig. 2(a) – (d). We found that this model excellently describes the magnetotransport results in the non-magnetic $In_{1-x}Ga_xAs$ channels with MPE.

$$\frac{\Delta\rho}{\rho} = -a^2 \ln\left(1 + b^2 H^2\right) + \frac{c^2 H^2}{1 + d^2 H^2}, \quad (1)$$

where

$$a = A_1 J D(\epsilon_F) \left[S(S+1) + \langle M^2 \rangle\right], \quad (2)$$

$$b^2 = \left[1 + 4S^2\pi^2\left(\frac{2JD(\epsilon_F)}{g}\right)^4\right]\left(\frac{g\mu_B}{\alpha k_B T}\right)^2, \quad (3)$$

$$c^2 = \frac{\sigma_1\sigma_2(\mu_1 + \mu_2)^2}{(\sigma_1 + \sigma_2)^2}, \quad (4)$$

$$d^2 = \frac{(\sigma_1\mu_2 - \sigma_2\mu_1)^2}{(\sigma_1 + \sigma_2)^2}. \quad (5)$$

In eq. (2) and (3), $A_1$ is a constant representing the contribution of spin scattering to the whole MR, $\alpha$ is a numerical factor that is on the order of unity, $D(E_F)$ is the density of states (DOS) at the Fermi level $E_F$, $g$ is the effective Lande factor of the $In_{1-x}Ga_xAs$ QW, $\langle M^2 \rangle$ is the averaged squared magnetization, $S$ is the localized spin moment of (Ga,Fe)Sb (we assume $S$ = 5/2 for $Fe^{3+}$ ions in (Ga,Fe)Sb), and $J$ is the s,p-d exchange interaction energy at the NM/FM bilayer interface. In eq. (4) and (5), $\sigma_i$ and $\mu_i$ represent the conductivity and mobility of electron carriers in $In_{1-x}Ga_xAs$, respectively. The subscripts 1 and 2 of each parameter denote the majority and minority spins, respectively. The first term on the right side of eq. (1), which gives a negative MR component is due to the Kondo scattering of electron carriers in the $In_{1-x}Ga_xAs$ by the localized Fe spins at the (Ga,Fe)Sb interface. The second term on the right side of eq. (1), which gives a positive MR component, is caused by the *s-d* exchange interaction between transport carriers in $In_{1-x}Ga_xAs$ and the localized Fe spins at the interface of $In_{1-x}Ga_xAs$/(Ga,Fe)Sb. The fitting results using eq.(1) are shown by black dashed curves in Fig. 2 (a)-(d), showing excellent agreement with the experimental MR curves at various $V_g$.

From the modified Khosla-Fischer model of eq. (1), the interesting behavior of the PMR in our samples can be understood as follows: The negative MR component, which is induced by the Kondo effect, is enhanced under large penetration *P* of the electron carrier wavefunction into the FMS (Ga,Fe)Sb, as we experimentally proved in the previous work [19]. Considering the FET operation with a two-dimensional electron gas channel in the $In_{1-x}Ga_xAs$ QW, there are three different regimes of the PMR depending on the gate voltage $V_g$, as shown in Fig. 3(b)-(d): (i, accumulation regime) As shown in Fig. 3(b), when a positive or zero $V_g$ is applied, the wavefunction is tilted towards the $Al_2O_3$ layer side and the penetration of the electron wavefunction into the (Ga,Fe)Sb side is small but not zero. Therefore, the PMR shows small negative MR. In this regime, the electron carrier concentration *n* is larger than other regimes. (ii, depletion regime) As shown in Fig. 3(c), when a small negative or zero $V_g$ is applied, the carrier wavefunction in the $In_{1-x}Ga_xAs$ QW is pushed towards the FM (Ga,Fe)Sb side, which leads to the larger negative MR. The electron carrier concentration *n* is smaller than that in (i). (iii, inversion regime) As shown in Fig. 3(d), when a large negative $V_g$ is applied, the channel is eventually inverted and hole carriers are induced at the $Al_2O_3$/InAs interface far from the FMS side. These hole carriers do not interact with the magnetic moments in the (Ga,Fe)Sb layer, and thus MPE is again suppressed. Also, our calculation suggests that there are holes accumulated in the (Ga,Fe)Sb layer near the InAs/(Ga,Fe)Sb interface, which may lead to non-linear magnetic field dependence in the Hall resistance (See Supplementary Material). As the regime goes close to (iii), the PMR should be decreased. Therefore, when the $V_g$ becomes more negative to enhance the penetration, the electrons are fully depleted eventually at negative large $V_g$ and there is no MPE at all. In other words, appropriate $V_g$ should be chosen to obtain the largest spin splitting $\Delta E$.

Next, we discuss the dependence of $\Delta E$ on $n$ and $\tau$ (= $\mu_n m^*/e$, $\mu_n$ is electron mobility), the data of which are shown in Fig. 4(a) and (b), respectively. $\Delta E$ in a two-dimensional electron system is estimated from the modified Khosla-Fischer model [19,29]:

$$\Delta E = 2\pi\hbar^2 \frac{n}{\mu_n m^*} d = \frac{2\pi\hbar^2}{e} \frac{n}{\tau} d, \tag{6}$$

where $d$ is the fitting parameter from eq. (5). $n$ and $\mu_n$ are estimated from the ordinary Hall effect measurements (Detailed discussions on the Hall effect data and analyses on In$_{1-x}$Ga$_x$As/(Ga,Fe)Sb are described in Supplemental Material). Here $m^*$ of In$_{1-x}$Ga$_x$As, $m^*_{\text{InGaAs}}$, is determined by a linear intrapolation between the effective mass of InAs ($m_{\text{InAs}}$) and GaAs ($m_{\text{GaAs}}$), taken from literature: $m^*_{\text{In}_{1-x}\text{Ga}_x\text{As}}(x,n) = m_{\text{InAs/(Ga,Fe)Sb}} + x[m_{\text{GaAs}}(n) - m_{\text{InAs}}(n)]$ [22,30], where $m_{\text{InAs/(Ga,Fe)Sb}}$ is obtained by our recent study of Shubnikov-de Haas (SdH) oscillations in InAs/(Ga,Fe)Sb [31]. We have recently found that $m_{\text{InAs/(Ga,Fe)Sb}}$ does not strongly depend on $n$, which is different from that in a NM InAs QW. In a usual NM InAs, $m^*$ increases with increasing $n$ because of the non-parabolicity of the InAs conduction band. However, in the present case of InAs/(Ga,Fe)Sb bilayers, when we apply a positive gate voltage $V_g$, $n$ is increased (making $m^*$ heavier) but $P$ is decreased (making $m^*$ lighter due to the smaller $s$-$d$ exchange interaction at the NM/FM interface). Thus, it is reasonable that $m^*$ only weakly depends on $n$ as observed experimentally [31]. According to eq (6), this result suggests that $n$ and $\tau$ significantly affect the spin splitting $\Delta E$.

Figure 4(a) shows the $n$-dependence of $\Delta E$ in all the FET devices (D$_0$, D$_5$, D$_{7.5}$, and D$_{10}$; solid circles), comparing with the previous result in InAs/(Ga,Fe)Sb ($\widetilde{D}_0$; open circles) [19]. Note that $\widetilde{D}_0$ has a same semiconductor heterostructure as D$_0$, and we re-estimate $\Delta E$ of the previous study in $\widetilde{D}_0$ [19] by taking into account the the $n$-dependence of $m^*$. At first, we found that unlike device $\widetilde{D}_0$ in the previous work, $\Delta E$ of device D$_0$, D$_5$, D$_{7.5}$, and D$_{10}$ increases with increasing $n$. In two particular devices $\widetilde{D}_0$ and D$_0$, both contain only an InAs channel ($x = 0$), the $\Delta E - n$ relationship shows a U-shape curve, where $\Delta E$ first decreases at $n < 1\times10^{13}$ cm$^{-2}$ (in device $\widetilde{D}_0$) but increases at higher $n > 2\times10^{13}$ cm$^{-2}$ (in device D$_0$) as $n$ increases. These results clearly indicate a general trend that the MPE is enhanced at larger $n$ when increasing $n$ above a specific threshold value. This can be explained by two counteracting effects of the gate voltage $V_g$ on the MPE in our FET devices; the electron wavefunction penetration ($P$)- and electron concentration ($n$)-induced effects. In the region with small $n$ (corresponding to regime (ii) at negative $V_g$), $P$ is large and the penetration-induced effect dominates the MPE. An increase of $V_g$ towards positive side decreases $P$ and thus suppresses MPE. At large positive $V_g$ (corresponding to regime (i)), while the penetration $P$ is minimized, a larger $n$ enhances the $s$-$d$ exchange interaction at the InAs/(Ga,Fe)Sb interface, as commonly observed in carrier-induced FMSs. Thus the carrier-induced effect becomes more prominent, leading to stronger MPE and a larger spin splitting at the large-$n$ region. These results clarify the important roles of both factors, the electron concentration $n$ and the penetration $P$ of the electron wavefunction, for increasing the MPE and $\Delta E$.

We note that the $\Delta E - n$ relationships in devices $D_0$, $D_5$, $D_{7.5}$, and $D_{10}$ do not show the increase of $\Delta E$ in the low-$n$ region whereas device $\widetilde{D}_0$ shows the increase of $\Delta E$ in the low $n$ region due to the $P$-induced MPE. Comparing with the case of device $\widetilde{D}_0$, all the FET channels in device $D_0$, $D_5$, $D_{7.5}$, and $D_{10}$ quickly reach the inversion regime (the abovementioned regime iii) at negative $V_g$, where holes are increased and parallel conduction occurs (see also Supplemental Material). The difference might originate from the weaker pinning effect of the Fermi level in the $Al_2O_3$/$In_{1-x}Ga_xAs$/(Ga,Fe)Sb FETs than that in the $HfO_2$/InAs/(Ga,Fe)Sb FETs in the previous work [19]. The Khosla-Fischer model is thus unable to be used to estimate $\Delta E$ in this low-$n$ region in devices $D_0$, $D_5$, $D_{7.5}$, and $D_{10}$. Further studies are required to determine $\Delta E$ in the NM channel in the low-$n$ region without using Hall resistance data. For example, observing the spin-split Fermi surface using SdH oscillations is a promising approach, which will be reported elsewhere [31].

Finally, we comment on the $\Delta E - \tau$ relationship shown in Fig. 4(b). Comparing all the data shown in Fig. 4(b), it is found that $\Delta E$ increases with decreasing $\tau$. From the modified Khosla-Fischer model, larger penetration $P$ of the carrier wavefunction of the NM channel into the (Ga,Fe)Sb side, which enhances $\Delta E$, leads to the larger spin-dependent scattering for the carriers and results in smaller $\tau$. Therefore, the negative correlation between $\Delta E$ and $\tau$ can be reasonably understood.

## IV. CONCLUSION

In conclusion, we investigate the MPE in semiconductor NM/FM heterostructures by characterizing the magnetotransport properties of In$_{1-x}$Ga$_x$As/(Ga,Fe)Sb heterostructures. We performed the systematic analysis of the MPE by measuring the gate-controlled PMR using standard FET operation. These results clearly indicate a general trend that the $\Delta E - n$ characteristics follow a U-shape, and the MPE is enhanced at larger $n$ when increasing $n$ above a specific threshold value. From the $\Delta E - n$ characteristics, there are two counteracting effects that affect the $\Delta E$; one is determined by the electron concentration which leads to carrier-induced $s$-$d$ exchange interaction at the NM/FM interface, and the other is determined by the penetration depth of the electron wavefunction into the ferromagnetic side. Therefore, in order to obtain large MPE, to choose the appropriate carrier concentration is crucial. The data obtained in this study will play an important role to unveil the microscopic mechanism of MPE in semiconductor-based non-magnetic/ferromagnetic heterostructures.

**ACKNOWLEDGEMENTS**

This work was partly supported by Grants-in-Aid for Scientific Research (18H05345, 20H05650, 20K15163, 20H02196), the CREST (JPMJCR1777) and PRESTO (JPMJPR19LB) Programs of JST, and the Spintronics Research Network of Japan (Spin-RNJ). A part of this work was conducted at the Advanced Characterization Nanotechnology Platform of the University of Tokyo, supported by the "Nanotechnology Platform" of the Ministry of Education, Culture, Sports, Science and Technology (MEXT), Japan.

**DATA AVAILABILITY**

The data that support the findings of this study are available from the corresponding author upon reasonable request.

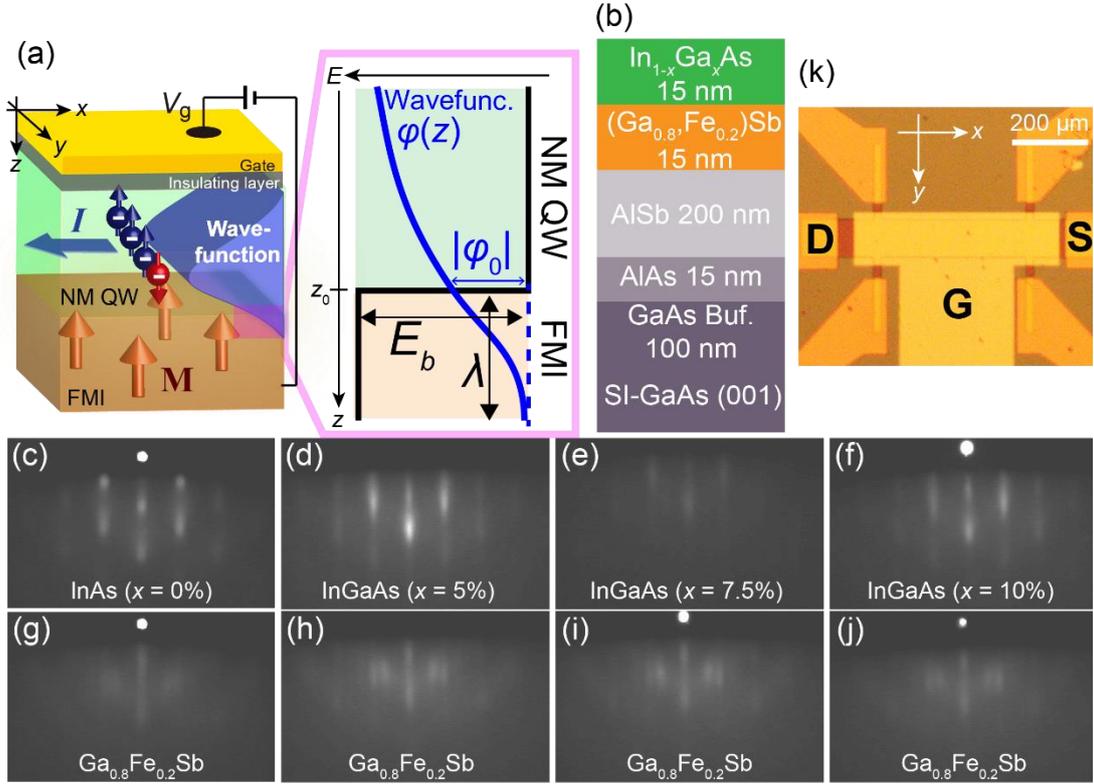

FIG. 1 (a) Schematic (left) device structure and (right) band alignment with electron carrier wavefunction (blue curve) in a NM QW/ FM insulator (FMI) heterostructure. In the left image, the wavefunction in the NM QW penetrates into the FMI. When an electrical current $I$ flows in the in-plane direction, electrons interact with the neighboring magnetization **M** at the interface, and are partially magnetized by MPE. A semiconductor-based NM/FM system enables modulation of the MPE by controlling the wavefunction with a top gate voltage $V_g$. In the right image, black solid line indicates the conduction band bottom of both materials. The carrier wavefunction penetrates into the FMI side with a penetration depth $\lambda$, determined by the barrier height $E_b$ and the wavefunction amplitude at the interface $|\varphi_0|$. (b) Sample structure of $D_0$, $D_5$, $D_{7.5}$, and $D_{10}$ with $x$ = 0%, 5%, 7.5% and 10%, respectively. (c)−(f) RHEED patterns of In$_{1-x}$Ga$_x$As and (g)−(j) those of (Ga,Fe)Sb during the MBE growth of In$_{1-x}$Ga$_x$As/(Ga,Fe)Sb heterostructures with $x$ = 0%, 5%, 7.5% and 10%, respectively. (k) Top-view optical microscopy image of the FET device ($x$ = 5%, $D_5$) examined in this study. We apply an electron current from the source (S) to the drain (D) and a gate voltage $V_g$ from the gate electrode (G) to S.

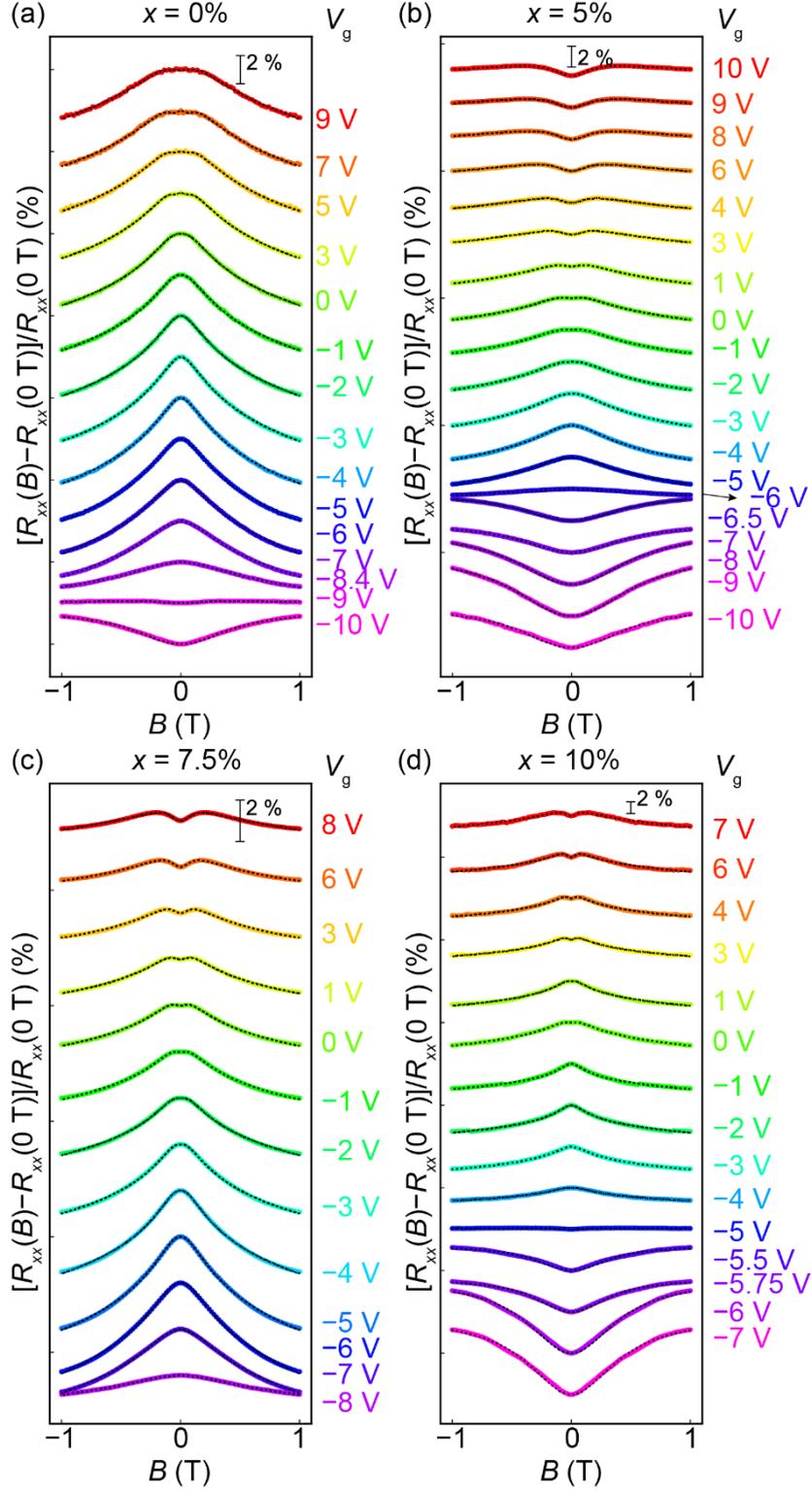

FIG. 2 (a)−(d) Magnetoresistance (MR) ($[R_{xx}(B)−R_{xx}(0\ T)]/R_{xx}(0\ T)$) of the In$_{1−x}$Ga$_x$As/(Ga,Fe)Sb heterostructures with $x = 0\%$, $5\%$, $7.5\%$ and $10\%$ under various $V_g$ with a perpendicular magnetic field $B$ (//[001] of GaAs substrate) at 3.5 K, respectively. The black dashed curves represent the fitting results from eq. (1). The MR curves of $x = 0\%$, $5\%$, $7.5\%$ and $10\%$ at each $V_g$ have been offset by 2.5%, 2.5%, 2% and 5% for a clearer view, respectively.

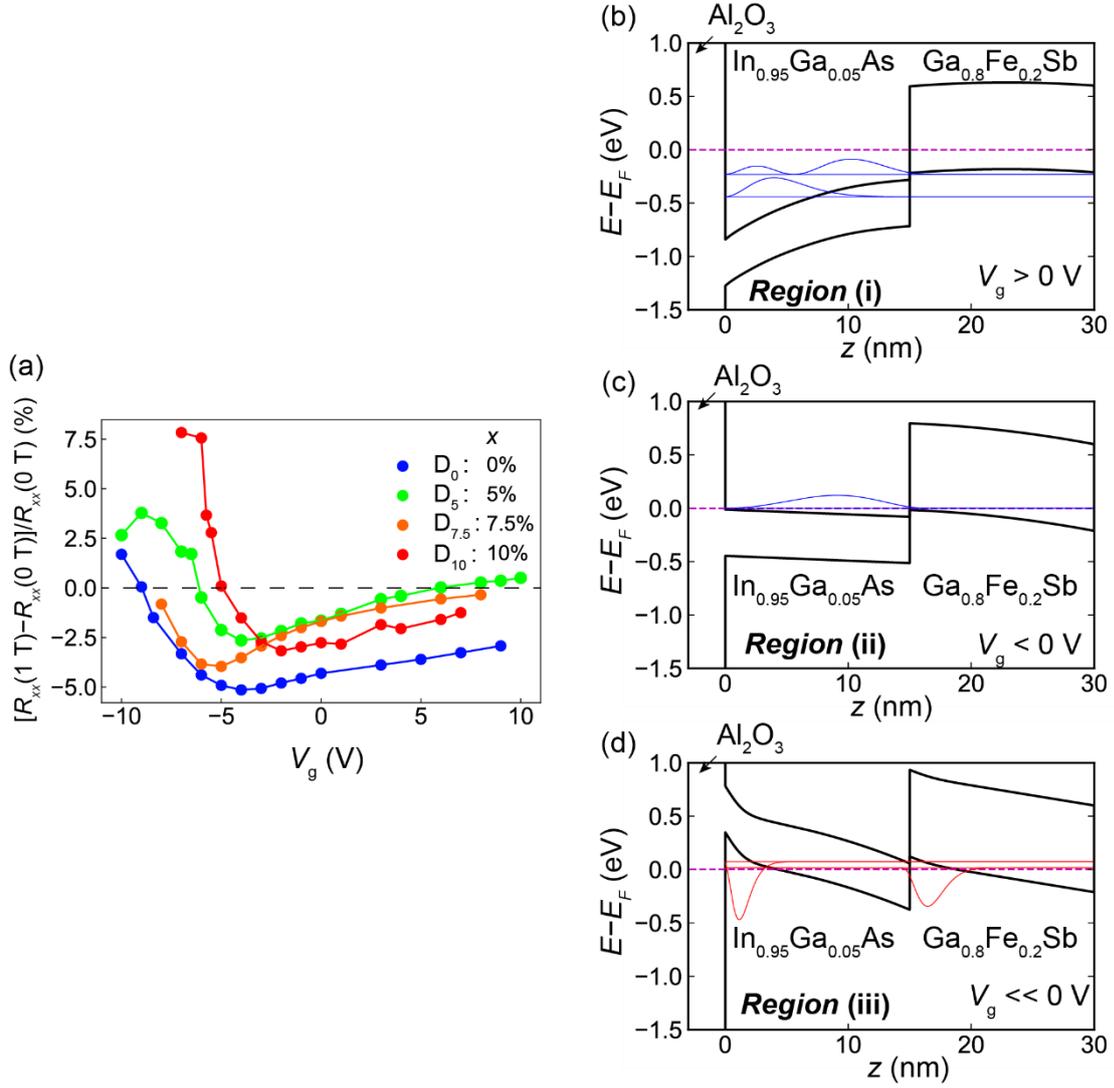

FIG. 3 $V_g$ dependence of the MR ratio at 1 T (= $[R_{xx}(1\text{ T})-R_{xx}(0\text{ T})]/R_{xx}(0\text{ T})$) in all devices examined in this work. Blue, green, orange, and red points indicate the data of device $D_0$, $D_5$, $D_{7.5}$, and $D_{10}$, respectively. (b) −(d) Self-consistent calculation results of the band profile in in device $D_5$ ($x = 5\%$) in the growth direction (// $z$). The blue and red curves indicate the electron and hole wavefunctions in accumulation ((b) $V_g > 0$ V, region (i)), depletion ((c) $V_g < 0$ V, region (ii)), and inversion ((d) $V_g << 0$ V, region (iii)) operation, respectively. Purple dashed line indicates the Fermi level $E_F$.

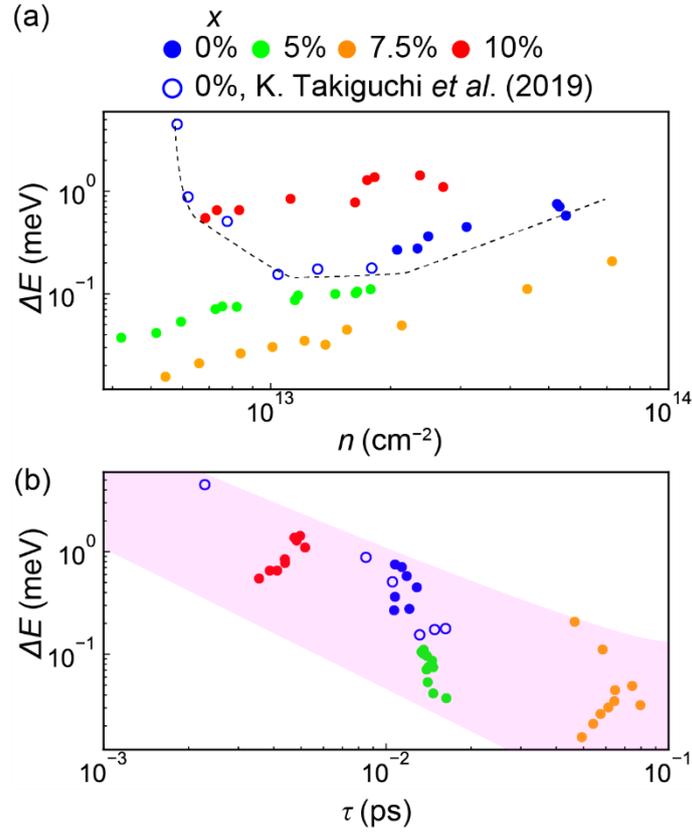

FIG. 4 (a) Spin splitting $\Delta E$ vs. electron concentration $n$ and (b) relaxation time $\tau$ of device $D_0$ (blue), $D_5$ (green), $D_{7.5}$ (orange), and $D_{10}$ (red). Blue open circles are for device $\widetilde{D}_0$ [19]. Black dashed line in (a) is an eye-guide for a clear view of the InAs/(Ga,Fe)Sb data. The pink shaded area in (b) represents the eye-guide for the negative correlation of $\Delta E$ vs. $\tau$.

# Supplemental Material

# Gate-controlled proximity magnetoresistance in $In_{1-x}Ga_xAs$/(Ga,Fe)Sb bilayer heterostructures

Kosuke Takiguchi[1], Kyosuke Okamura[1], Le Duc Anh[1,2,3], & Massaaki Tanaka[1,4,5]

[1] *Department of Electrical Engineering and Information Systems, The University of Tokyo, Bunkyo-ku, Tokyo 113-8656, Japan*

[2] *Institute of Engineering Innovation, The University of Tokyo, Bunkyo-ku, Tokyo 113-8656, Japan.*

[3] *PRESTO, Japan Science and Technology Agency, Kawaguchi, Saitama, 332-0012, Japan*

[4] *Centre for Spintronics Research Network, The University of Tokyo, Bunkyo-ku, Tokyo 113-8656, Japan.*

[5] *Institute for Nano Quantum Information Electronics, The University of Tokyo, 4-6-1, Komaba, Meguro-ku, Tokyo 153-8505 Japan.*

**Hall effect variation with gate voltage in In$_{1-x}$Ga$_x$As/(Ga,Fe)Sb**

As we discussed in the main manuscript, in order to estimate the spin-splitting energy $\Delta E$ from the PMR curve fitting, we need to measure the electron carrier concentration $n$ and mobility $\mu_n$. Figure S1(a)-(d) show the Hall measurement results of device D$_0$, D$_5$, D$_{7.5}$, and D$_{10}$, respectively. For the estimation of $n$ and $\mu_n$, we fit the $B$-linear component corresponding to the ordinary Hall effect (black dashed line) to the Hall resistance data in small negative $V_g$ ~ large positive $V_g$ regions, shown by the yellow-colored regions in Fig. S1(a)-(d). In these regions, $n$ monotonically decreases with decreasing $V_g$, which is consistent with the standard FET operation from regime (i) to (ii), as explained in the main manuscript.

On the other hand, the Hall resistances in all the devices show non-linear behavior especially in the large negative $V_g$ regions, shown by the blue-colored regions in Fig. S1(a)-(d). In these regions, the non-linearity is too strong to be fit by the ordinary Hall effect. The non-linear Hall resistance curves in the In$_{1-x}$Ga$_x$As/(Ga,Fe)Sb bilayer heterostructures may be caused by two reasons, an anomalous Hall effect (AHE) and/or two-carrier conduction. In bilayer systems with MPE, it is reported that the Hall resistance shows the hysteresis curve corresponding to the magnetization of the FM layer even if the transport carriers mainly flow in the NM channel [1,2]. However, if this is the case, the Hall resistance vs. $B$ curve should be proportional to the magnetization of the ferromagnetic (Ga,Fe)Sb layer. We do not observe any coercivity in the Hall measurements which corresponds to that of the (Ga,Fe)Sb magnetization. Thus, we consider that the non-linearity observed in our devices originates from the two-carrier conduction. We note that, as mentioned in the main manuscript, the modified Khosla-Fischer model is not applicable to the two-carrier regions. Thus, we do not present any data of the spin splitting in Fig. 4.

In order to estimate the carrier density and mobility in two-carrier conduction region, we fit a two-carrier model [3] to the experimental Hall resistance $R_{xy}$ – magnetic field $B$ curves, as follows:

$$R_{xy} = \frac{p\mu_p^2 - n\mu_n^2 + (\mu_n\mu_p B)^2 (p-n)}{e\left[(p\mu_p + n\mu_n)^2 + (\mu_n\mu_p B)^2 (p-n)^2\right]} B, \tag{S1}$$

where $p(n)$, and $\mu_{p(n)}$ are the carrier concentration and mobility of holes (electrons), respectively. At various $V_g$, the fitting curves using eq. (S1) are shown by yellow solid curves in the blue-colored regions of Fig. S1 (a)-(d). Figure S2 shows the fitting results. In the large negative $V_g$ region (blue region), there are extra holes with relatively larger $\mu_p$ than $\mu_n$. There are two possible origins of this high-mobility holes: hole carriers in the InGaAs inversion layer and those in the valence band of (Ga,Fe)Sb. The holes in the InGaAs inversion layer is located far from the interface of (Ga,Fe)Sb as shown in Fig. 3(d), which do not interact via the s-d exchange coupling with the Fe spins in the FMS. Thus, they can possess high $\mu_p$. Meanwhile, the holes in the coherent valence band of (Ga,Fe)Sb can also possess $\mu_p$ that is much higher than that ($< 10$ cm$^2$/Vs) of the holes in the Fe-related impurity band [4] [5]. As shown in Fig. 3(d), these two kinds of holes can exist in the strong negative $V_g$ region. We notice that, the trend of $n$ and $p$ against $V_g$ does not match the standard FET operation corresponding to regime (iii). Since we do not have a clear answer for the carrier distribution and origin of holes, further study will be required.

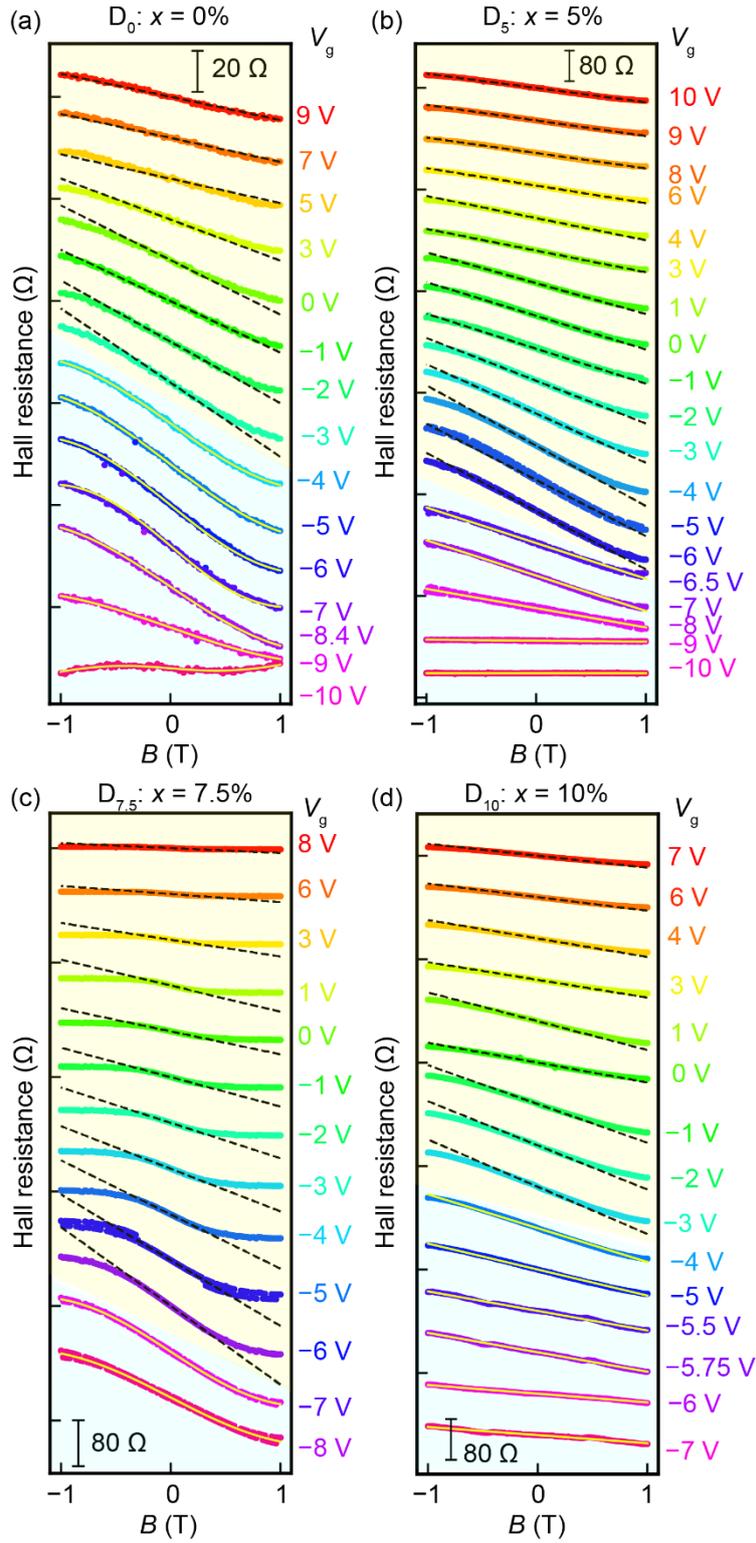

Figure S1 (a)-(d) Hall resistance curves of of $x = 0\%$, 5%, 7.5% and 10% under various $V_g$ with perpendicular magnetic field $B$ (//[001] of GaAs substrate) at 3.5 K, respectively. Black dashed lines in the yellow regions represent the fitting results of the ordinary Hall effect. Yellow solid curves in the blue regions represent the fitting results using eq. (S1). The Hall resistance curves at various $V_g$ have been offset by 20 Ω, 80 Ω, 80 Ω and 80 Ω for easy viewing.

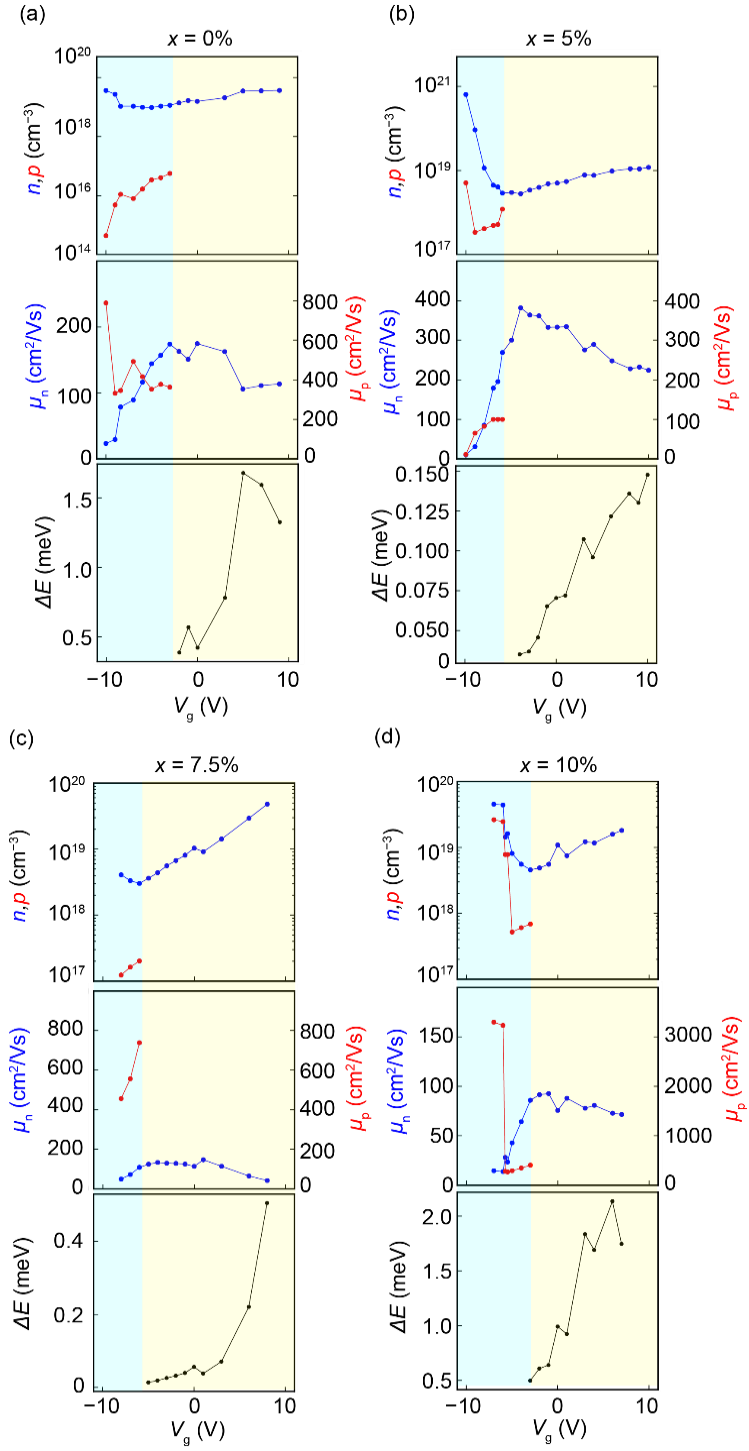

Figure S2 Fitting results of $n$, $p$, $\mu_n$, $\mu_p$, and $\Delta E$, as a function of $V_g$ with various $x$: (a) $x = 0\%$, (b) $x = 5\%$, (c) $x = 7.5\%$ and (d) $x = 10\%$, (upper panel) carrier concentration $n(p)$ and (middle panel) mobility $\mu_{n(p)}$ of electrons (holes) $vs.$ $V_g$ obtained by the fitting shown in Fig. S1. The blue and red points indicate electrons' and holes' parameters, respectively. (lower panel) Spin splitting $\Delta E$ $vs.$ $V_g$. In the blue and yellow regions, the parameters are obtained by the two-carrier model and ordinary Hall effect, respectively. In the blue regions, $\Delta E$ cannot be measured since the modified Khosla-Fischer model is not applicable to the two-carrier conduction system.